\documentclass[pra, superscriptaddress, twocolumn]{revtex4}

\usepackage{amsfonts}
\usepackage{amssymb}
\usepackage[dvips]{graphicx}
\usepackage{amsmath}

\newcommand{\ket}[1]    {| #1 \rangle}
\newcommand{\braket}[2]{\langle #1 | #2 \rangle }

\begin{document}
\title{Quantum states made to measure}

\author{Konrad Banaszek}
\affiliation{Institute of Theoretical Physics, University of Warsaw, ul. Ho\.{z}a 69, PL-00-681 Warszawa, Poland}
\affiliation{Institute of Physics, Nicolaus Copernicus University, ul.~Grudziadzka 5, PL-87-100 Toru\'{n}, Poland}
\author{Rafa{\l} Demkowicz-Dobrza{\'n}ski}
\affiliation{Institute of Physics, Nicolaus Copernicus University, ul.~Grudziadzka 5, PL-87-100 Toru\'{n}, Poland}
\author{Ian A.~Walmsley}
\affiliation{Clarendon Laboratory, University of Oxford, Parks Road, Oxford OX1 3PU, United Kingdom}

\date{\today}
\begin{abstract}
Recent progress in manipulating quantum states of light and matter brings quantum-enhanced measurements closer to prospective applications. The current challenge is to make quantum metrologic strategies robust against imperfections.
\end{abstract}
\maketitle

Precision measurements lie at the heart of modern science and engineering. In order to determine a parameter of an object---its position, speed, mass, frequency, phase etc.---a probe, such as a light beam or a collection of atoms, is made to interact with that object, in such a way that the state of the probe is altered depending on the value of the parameter of interest. Ultimately one must consider the probe to be quantum mechanical, which limits the sensitivity of the arrangement: as a rule the final quantum states of the probe cannot be distinguished perfectly, even in principle. Proper understanding of the limits to precision set by this indeterminacy requires consideration of what resources are employed in building the probe and what imperfections, such as loss of the probe particles or noise leading to random corruption of the probe state, are present in the device. Quantum optics is making impressive advances in exploring these questions.

At the most basic level, a light beam employed as a probe can be treated as a classical entity with well defined properties. But quantum indeterminacy will manifest itself at the detection stage in the statistical character of photocount events. Within the fully quantum framework one has much broader possibilities to manipulate the state of optical radiation. Two specific features that are exploited in quantum-enhanced metrology are the ability to prepare states with lower intrinsic noise than their classical counterparts, and the existence of states that exhibit stronger correlations between individual systems than it is allowed classically.

\begin{figure*}
\includegraphics[width=0.85 \textwidth]{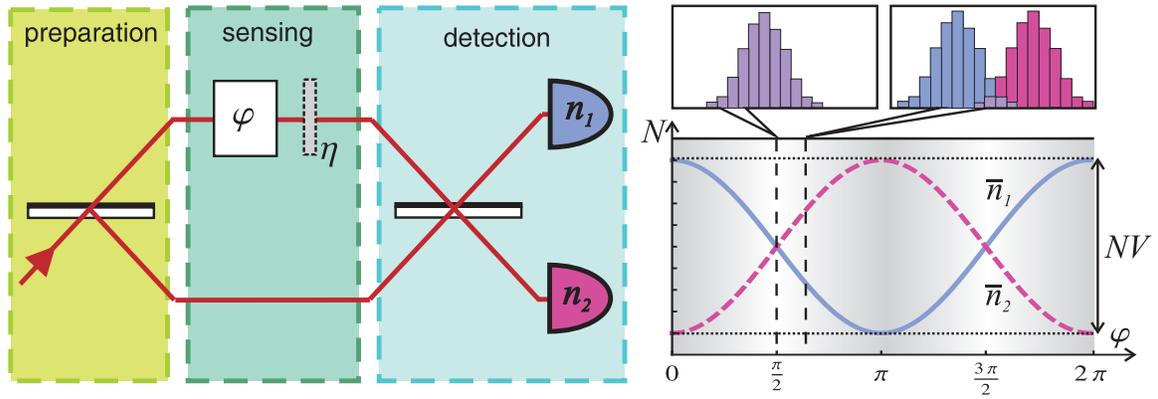}
\caption{A Mach-Zehnder interferometer. The phase shifter $\varphi$ inserted in the upper arm modulates the intensities at the output of the interferometer. The detectors produce photocount numbers $n_1$ and $n_2$ whose expectation values $\bar{n}_1$ and $\bar{n}_2$ exhibit a fringe pattern shown in the right panel. The visibility $V$ characterizes the relative modulation depth of interference fringes and $N=\bar{n}_1+\bar{n}_2$ is the total average number of photocounts. Possible linear losses accompanying the phase shift are represented by an attenuator with intensity transmission $\eta$.}
\label{fig:mz}
\end{figure*}

An insightful illustration of these issues, relevant to a number of practical sensing schemes, is provided by an optical phase measurement shown in Fig.~\ref{fig:mz}. A light beam is sent into a Mach-Zehnder interferometer to sense a phase $\varphi$. The detectors monitoring output beams produce random numbers of photocounts which are used to guess the value of $\varphi$. If our task is to determine a small phase shift around a fixed operation point, the minimum achievable measurement uncertainty is $1/\sqrt{N}$ where $N$ is the average total number of photocounts. This is the well established {\em shot noise limit}, whose scaling with $N$ can be viewed as the reduction in statistical uncertainty obtained by repeating an elementary observation $N$ times.

\begin{figure}
\includegraphics[width=1 \columnwidth]{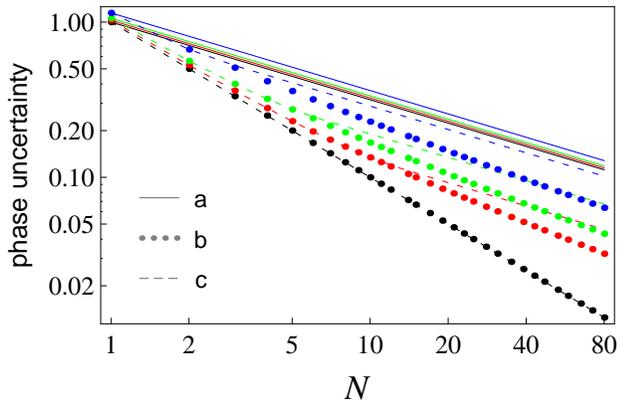}
\caption{The log-log plot of the measurement uncertainty for the probe arm transmission $\eta$ equal to $100\%$ (black), 90\% (red), 80\% (green), and 60\% (blue). {\bf a}, Solid lines depict the shot-noise-limited operation of an interferometer optimized for a given value of losses as a function of the average number of photocounts $N$.
{\bf b}, Dots represent precision that can be achieved with an engineered optimal $N$-photon state following the approach described in \protect\cite{DornDemkPRL09}. {\bf c}, Dashed lines indicate the lowest uncertainty that can be achieved with single photons in a simple multipass strategy, with $N$ being the product of the number of photons used and the number of passes for each photon.}
\label{fig:scaling}
\end{figure}

A common deleterious optical effect is linear loss, which typically cannot be separated from the phase shift of interest. Intuitively, attenuation needs to be compensated by directing a higher fraction of input light towards the sensing arm. Rigorously, the optimal setting can be found by maximization of the Fisher information \cite{Fisher}, which quantifies how much information about a small change in a parameter of interest can be obtained from random variables whose statistics depends conditionally on that parameter. Interestingly, in contrast with the lossless case the fringe visibility at the output of the interferometer is no longer 100\%, as one needs to strike a balance between the modulation depth and the total number of photocounts. This feature emphasizes the importance of selecting the right performance criterion. Quantitatively, attenuation lowers the proportionality factor in the $1/\sqrt{N}$ scaling, as shown in Fig.~\ref{fig:scaling}a.

Throughout the above discussion, the light beams were described as classical fields with well defined properties. In the fully quantum picture, each photocount is triggered by an absorption of a photon from the incident beam. If all the photons are injected into one input port of the interferometer, we can describe their evolution individually following the remark of P. A. M. Dirac that each photon ``interferes only with itself'' \cite{Dirac}. In the lossless case, the first beam splitter prepares a single photon in a superposition state $\frac{1}{\sqrt{2}}(\ket{10} + \ket{01})$, where the two digits within kets specify the number of photons in the upper and the lower interferometer arm. The phase shifter alters the state to $\ket{\psi(\varphi)} = \frac{1}{\sqrt{2}}(e^{i\varphi}\ket{10} + \ket{01})$. For different phase shifts these states are generally non-orthogonal and therefore cannot be distinguished perfectly. The overlap between two states differing by a small phase shift $\delta\varphi$ equals approximately to $\bigl|\braket{\psi(\varphi)}{\psi(\varphi+\delta\varphi)}\bigr|^2
\approx 1-\frac{1}{4}(\delta\varphi)^2$. The task of discrimination becomes easier if we send $N$ identically prepared photons, each one of them emerging in a state either $\ket{\psi(\varphi)}$ or $\ket{\psi(\varphi+\delta\varphi)}$. The overlap scales exponentially with the number of copies, yielding $\bigl|\braket{\psi(\varphi)}{\psi(\varphi+\delta\varphi)}\bigr|^{2N} \approx 1-\frac{1}{4}(\sqrt{N}\delta\varphi)^2$. Thus the same distinguishability is obtained for a phase shift reduced by a factor $1/\sqrt{N}$. This is the quantum mechanical explanation of the shot noise limit \cite{CavePRL80}, revealing its
more fundamental roots than the mere characteristics of the detectors.

The first beam splitter 
of the interferometer can be viewed as the preparation stage of a certain quantum state to sense the phase shift. Leaving aside technical difficulties, one may wonder whether another state could offer better sensitivity. A more exotic, highly correlated superposition $\frac{1}{\sqrt{2}}(\ket{N0} + \ket{0N})$ of all $N$ photons travelling in either the upper or the lower arm---dubbed a {\em N00N state} \cite{BollItanPRA96,DowlPRA98}---after the phase shift takes
the form $\ket{\psi_N(\varphi)} = \frac{1}{\sqrt{2}} ( e^{iN\varphi}\ket{N0} + \ket{0N} )$,
as the phase acquired is proportional to the number of photons sent through the shifter. It is now easy to check that for a small phase shift $\bigl|\braket{\psi_N(\varphi)}{\psi_N(\varphi+\delta\varphi)}\bigr|^2 \approx 1-\frac{1}{4}(N\delta\varphi)^2$, which means that the measurement uncertainty scales as $1/N$ with the number of photons. This represents a substantial enhancement over the shot noise limit, and saturates the ultimate bound defined by quantum mechanics, known as the {\em Heisenberg limit} \cite{GiovLloySCI04}, for this sensing interaction and a given total number of photons.

In the presence of losses, all the $N$ photons in a N00N state must survive attenuation in order to maintain the phase sensitivity. For the probe arm transmission $\eta$, the probability that none of $N$ photons gets lost is $\eta^N$. This unfavourable scaling with $N$ yields precision worse than the shot noise limit unless attenuation is minute or a few-photon state is used \cite{RubiKausPRA07,GilbHamrJOSAB08}. One may consider a general class of quantum states and analyze how much information about $\varphi$ is left after the combined action of the phase shift and losses. The amount of information can be quantified with the quantum analog of Fisher information calculated for the quantum state after the sensing interaction. Full optimization over probe states reported recently \cite{DornDemkPRL09} yields results shown in Fig.~\ref{fig:scaling}b. A careful choice of the input state allows us to go beyond the shot noise limit even in the presence of losses, but numerics indicates that the precision scaling becomes generally worse than that of the Heisenberg limit.

The efficiency of a sensing scheme depends on how resources are used and counted. If one sends a single photon into a Mach-Zehnder interferometer and lets the probe beam pass $N$ times through a lossless phase shifter, the state becomes $\frac{1}{\sqrt{2}}(e^{i N \varphi} \ket{10}+ \ket{01})$ offering the same sensitivity as the N00N state. This Heisenberg-type scaling in the number of passes, achieved without engineered quantum states of light, was recently developed into more elaborate schemes \cite{HiggBerrNAT07}. However, it is unclear whether these benefits can be fully retained in the presence of losses, as illustrated with a simple multipass strategy in Fig.~\ref{fig:scaling}c.

Identifying optimal states is only a stepping stone towards practical realization of quantum-enhanced interferometry, leaving outstanding issues how to prepare optimal states and how to measure them after the sensing interaction. In order to engineer required quantum states, one needs to have sources of {\em non-classical light} \cite{DellDeSiPHR06} whose properties cannot be described on the grounds of the classical theory of electromagnetism. So far, the most popular choice is parametric down-conversion in nonlinear media producing highly correlated pairs of photons that can be manipulated using linear optics into a variety of quantum states. In particular, feeding a photon pair into two input ports of a Mach-Zehnder interferometer yields a two-photon N00N state producing double-density fringes  \cite{RariTapsPRL90}. Experimental progress enabled demonstrations of higher fringe densities obtained using three- \cite{MitcLundNAT04} and four-photon \cite{WaltPanNAT04} states, eventually reaching visibilities that beat the shot-noise limit \cite{NagaOkamSCI07}.

In most of these proof-of-principle works, events of interest are accompanied by a background involving fewer photons than required. This background is dismissed when processing coincidence counts. But fewer-photon cases should also be accounted as a resource consumed for sensing. Further, most preparation schemes have a non-unit success rate, removing some of the produced photons before the sensing stage. These issues open up questions about the overall efficiency of realizations of quantum enhancements, with similar considerations applying to the detection stage. Extensive effort is currently dedicated to develop manipulation tools for photonic states free from such problems \cite{MoslLundPRL08}. This is essential to further progress in many areas of quantum technologies besides quantum-enhanced metrology \cite{OBriSCI07}. An intriguing parallel question is to what degree imperfections in manipulations can be tolerated or mitigated to preserve quantum enhancements. Current results in quantum optical information processing \cite{BourEiblPRL05,LuGaoPNAS08} allow for optimism, although they cannot be directly transferred to metrologic protocols owing to different performance criteria.

\begin{figure}
\includegraphics[width=0.6 \columnwidth]{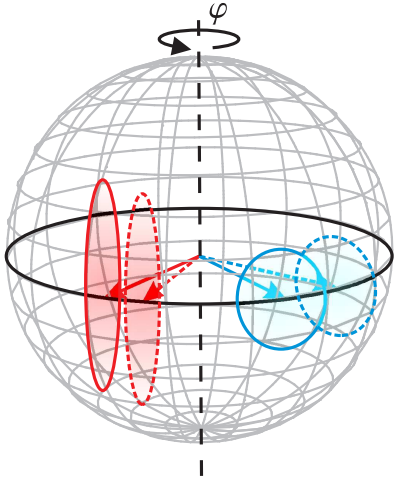}
\caption{A pictorial representation of the state of light in a Mach-Zehnder interferometer as a three-component real vector analogous to the Poincar\'{e} vector for the polarization state of light. A phase shift $\varphi$ corresponds to a rotation about the vertical axis. Quantum fluctuations make the tip of the vector spread out into a circular patch for shot-noise-limited interferometry (blue). Squeezing the uncertainty area in an appropriate direction results in enhanced phase sensitivity (red).}
\label{fig:poincare}
\end{figure}

There is an alternative way to look at precision limits of phase sensing, summarized in Fig.~\ref{fig:poincare}. Classically, the state of light inside the interferometer can be characterized using a three-component real vector. Its vertical orientation corresponds to entire light concentrated in the upper or the lower arm of the interferometer, while a phase shift induces a rotation about the vertical axis. To achieve phase sensitivity, the first beam splitter prepares a superposition located in the equatorial plane. In the quantum picture, the tip of the vector is spread out because of the Heisenberg uncertainty relation. For classical light sources, the spread is evenly distributed between horizontal and vertical directions. A natural thought is to reduce the horizontal uncertainty at the cost of the vertical one, leading to better distinguishability for a small phase shift. This is the basic idea behind {\em squeezed states} \cite{Squeezing} that can be currently generated with more than 10~dB noise reduction in power \cite{VahlMehmPRL08}. Further exploration of open pathways should eventually lead to robust and efficient quantum-enhanced metrology schemes.

The above considerations apply also to atomic systems, with two interferometer arms corresponding to different energy levels. Precision measurements of time and frequency can be carried out by sensing a relative phase shift using Ramsey interferometry. Analogously to the photonic case, independent preparation of individual atoms results in the shot noise limit that can be beaten by careful engineering of collective entangled states of the entire atomic ensemble \cite{EsteGrosNAT08,AppeWindPNAS09}. The task is even more challenging owing to the complex structure of real atoms and their environmental couplings \cite{YeKimbSCI08}, but this richness also creates new opportunities to control the sensing interaction leading to enhanced precision \cite{BoixDattPRL08}.

Quantum metrology has potential to impact both fundamental science, such as gravitational wave detection, as well as technology, for example frequency standards and spectroscopic sensing of molecules. But practical quantum-enhanced strategies must provide a clear benefit after weighing in extra complexities of manipulating quantum systems. To reach that stage more advanced quantum engineering techniques are needed. In parallel, theoretical foundations would gain a lot from a general framework for quantifying and comparing resources in metrologic schemes.

\end{document}